\documentclass[aps]{revtex4}
\usepackage{amsmath,amssymb}
\usepackage{color,graphicx}
\usepackage{multirow}
\usepackage{amsmath,amssymb}
\newcommand{\be}{\begin{equation}}
\newcommand{\ee}{\end{equation}}
\begin{document}
 \title{Fractional Laplacians  and L\'{e}vy flights in bounded domains}
 \author{Piotr Garbaczewski}
 \affiliation{Institute of Physics, University of Opole, 45-052 Opole, Poland}
 \date{\today }
 \begin{abstract}
  We  address   L\'{e}vy-stable  stochastic  processes in bounded domains, with a focus
   on a discrimination  between   inequivalent   proposals  for
    what  a boundary data-respecting  fractional  Laplacian  (and  thence the induced  random process)
     should actually be.  Versions  considered  are: restricted  Dirichlet,  spectral Dirichlet
  and   regional (censored)  fractional Laplacians. The affiliated random processes comprise:  killed,
   reflected and   conditioned  L\'{e}vy  flights, in particular those with an infinite life-time.
   The related concept of quasi-stationary distributions is briefly mentioned.
   \end{abstract}
 \maketitle

  \section{Motivation}

Jump-type   L\'{e}vy processes  in a bounded domain are  a subject of an active study  both in
  physics and mathematics  communities. The physics-oriented  research is conducted
  with  some disregard to an ample coverage  of  the   topic in the past and modern  mathematical literature.
    The reason is rooted  not only   in a   methodological gap between the   practitioners'  pragmatism   and
     the  mathematically rigorous  reasoning.  An important factor is a scarce (or even lack of) communication
   between   various research  groups and  research  streamlines. This refers not only to rather residual    physics-mathematics interplay,
   but also   to the mathematics community  per se:  relevant publications are  scattered in a large number of  highly specialized
     journals  and easily escape the attention  of potentially interested parties.

Recently an attempt has been  made  to establish a  common  conceptual  basis
for varied frameworks in which  fractional Laplacians appear.
Formally looking different,  but  actually  equivalent,  definitions
   of fractional Laplacians, appropriate for the description of L\'{e}vy stable processes
     in $R^n$, $n\geq 1$, have been collected and their   mutual  relationships analyzed in minute
     detail  in Ref.  \cite{kwasnicki}.

There is a general consensus that  the  standard Fourier multiplier
definition  appears  to be defective,   if  one passes to L\'{e}vy
flights  in a bounded domain.  This  is a consequence of  an inherent nonlocality of L\'{e}vy-stable  generators.
 Different proposals for boundary-data-respecting fractional Laplacians were given in the
literature.  Often,  with a view towards more efficient computer-assisted calculations (that mostly in connection with
nonlinear fractional differential equations, \cite{vazquez,bonforte}).

  However,  in contrast to the situation in $R^n$,  these proposals  are  known to be  inequivalent,  c.f.
   Refs. \cite{servadei}-\cite{zaba1}, see also \cite{bogdan,reflected}. Likewise, the   induced  jump-type processes are inequivalent
    and have different statistical characteristics.  That in particular refers to  a standard physical
     inventory, adapted  directly  from the  Brownian motion studies  \cite{redner}:
   the statistics of exits  from the domain,  e.g. first and mean first  exit times,
      probability of survival, its long time behavior  an ultimate  asymptotic decay
       \cite{gitterman}-\cite{dybiec1}.

 Interestingly, the existence problem for  jump-type processes  with an  infinite life-time in a bounded
 domain, seems to  have been left aside  in the physics  literature  (compare e.g. Ref. \cite{killing} in
  connection with  diffusion  processes and  Ref. \cite{acta} for  a preliminary discussion of the
  Cauchy process    in the interval).   To the contrary,   permanently trapped   L\'{e}vy-type  processes
   (diffusion processes like-wise)   have their place  in the mathematical literature.

    One category  of such processes  stems from the analysis of the
   long-time behavior of the survival probability   in case of absorbing  enclosures which actually allows to
single out  appropriate  conditioned  processes   that  never leave the domain once started within.
A related topic is that of quasi-stationary distributions (c.f. \cite{collet} in the random walk and
 Brownian contexts) and  the concept of so-called Yaglom limits, \cite{bogdan1,palmowski}.

   Another category refers to reflecting boundary data  and  to  reflected   L\'{e}vy-stable
    processes (the reflected Brownian motion  might be invoked at this point and set against the
    killed/absorbed one,  c.f. \cite{gitterman,bickel0,bickel}).  Actually,  censored fractional
      Laplacians  are interpreted as generators of reflected L\'{e}vy stable processes,
      \cite{bogdan,reflected}.

Let us concisely state the main problem addressed.   While giving meaning to the Laplacian in a bounded domain  $D\subset R^n$,
 denoted tentatively $\Delta _D$, we must account for various admissible boundary data,  that are  local i.e.  set on  the boundary $\partial D$  of an  open set $D$.
One may try to  define a fractional power of the Laplacian by importing its  locally defined   boundary data on $\partial D$,   through so-called
 spectral  definition  $(- \Delta _D)^{\alpha /2}$.

This  operator   is known to be different from   the   outcome  of the  procedure  in which one first executes  the fractional power of the Laplacian,
and next imposes  the boundary data, as embodied in the notation $(- \Delta )^{\alpha /2}_D$.   In case of absorbing boundaries,
 in contrast to $(- \Delta _D)^{\alpha /2}$,  the Dirichlet  boundary data
for   $(- \Delta )^{\alpha /2}_D$ need to be imposed as exterior  ones i.e. in the  whole  complement $R^n\backslash D$ of $D$.

Notwithstanding,  reflected Brownian motions belong to the bounded domain paradigm, \cite{bickel0,bickel} and  the  related  issue of reflected
L\'{e}vy flights should be explored,  that  in conjunction with the concept of censored  and /or regional fractional  Laplacians, \cite{bogdan,reflected}.

\section{Generalities}
\subsection{Transition densities}

Let us restate  our motivations  in a more formal lore  (our notation is consistent with that in Ref. \cite{lorinczi}).
  Namely, given the  (negative-definite) motion  generator $L$,  we shall
 consider  the (contractive)  semigroup  evolutions   of the form
\be
f(x,t) = T_tf(x)=    (\exp (t L)]f)(x)  = \int _{R^n} k(x,0;y,t) f(y) dy   =  E^x [f(X_t)], \ee
 where  $t\geq 0$.
In passing, we have  here  defined a local expectation value  $E^x[...]$, interpreted as  an average   taken at time  $t>0$,
  with  respect  to the process $X_t$ started in $x$ at $t=0$,  with values  $X_t=y \in R^n$  that are  distributed
  according to the  positive  transition  (probability)  density  function  $k(x,0;,y,t)$.

  We in fact deal with a bit more  general transition function
     $k(x,s;y,t), 0\leq s<t$ that is symmetric with respect to $x$ and $y$, and  time homogeneous. This  justifies
  the  notation $k(x,s;y,t)= k(t-s,x,y)=k(t-s,y,x)$ and subsequently  $k(x,0;y,t)=  k(t,x,y)=k(t,y,x)$.
  The "heat" equation
\be
\partial _t f(x,t)= L f(x,t)
\ee
 for $t \geq 0$  is    here   presumed to  follow.  We recall, that given a  suitable  transition function, we
 recover the semigroup   generator via
 \be
  [Lf](x)= \lim_{t\rightarrow 0} {\frac{1}{t}}  \int_{R^n}
 [k(t,x,y) f(y)dy - f(x)].
 \ee
in accordance with an (implicit strong continuity) assumption that  actually  $T_t=\exp(Lt)$.

For completeness let us mention that the semigroup property $T_t  T_s =T_{t+s}$,  implies
the validity of  the composition rule    $\int_{R^n} k((t,x,y)\, k(s,y,z)\, dy = k(t+s,x,z)$.

Let $B\subset R^n$, a probability that a subset $B$ has been reached by the process  $X_t$   started  in  $x\in R^n$,   after the
 time lapse  $t$,  can be inferred from   $ P[X_t \in B|X_s=x]= \int_B k(t-s,x,y)dy$, $0\leq s<t$, and reads
\be
 P^x(X_t\in B) = \int_B k(t,x,y)dy = k(t,x,B)
\ee
Clearly, $P^x(X_t\in R^n)=1$.

In general,  for time-homogeneous processes,   we have  $k(x,s;B,t) = \int _B k(t-s,x,y) dy$,   $s<t$,
hence  we can rephrase the Chapman-Kolmogorov relation as  follows:
 \be
 \int_{R^n} k(x,s;z,u) k(z,u,B,t) dz = k(x,s;B,t)= k(t-s,x,B)=  P[X_{t-s} \in B|X_s=x],
 \ee
 where $s<u<t$.

\subsection{Absorbing boundaries and survival probability}

Now,  we shall pass  to killed Brownian and L\'{e}vy-stable motions in a bounded domain.
Let us denote $D$ a bounded open set in $R^n$. By $T_t^D$ we denote  the semigroup  given by the process $X_t$ that is killed on exiting $D$.
Let  $k_D(t,x,y)$ be the transition density for $T_t^D$. Then  \cite{kulczycki}:
\be
T_t^Df(x)= E^x[f(X_t); t <\tau _D] =  \int_D  k_D(t,x,y) f(y) dy
\ee
provided  $x\in D$ , $t>0$ and   the first exit time $\tau _D= \inf \{t \geq 0, X_t \notin D \}$  actually stands for the  killing time  for $X_t$.

From the general theory of killed semigroups in a bounded domain there follows that in  $L^2(D)$ there exists  an  orthonormal    basis of eigenfunctions
$\{ \phi _n \}, n=1,2,...$ of $T_t^D$ and corresponding eigenvalues $\{ \lambda _n, n=1,2,... \}$  satisfying   $0<\lambda _1 <\lambda _2 \leq \lambda _3 \leq ...$.
Accordingly there holds $T_t^D \phi _n(x) = e^{-\lambda_nt}\, \phi _n(x)$, where  $x\in D, t>0$ and we also have:
\be
k_D(t,x,y) = \sum_{n=1}^{\infty } e^{-\lambda_nt}\, \phi _n(x)\, \phi _n(y)
\ee
The eigenvalue $\lambda _1$ is non-degenerate (e.g. simple)  and the corresponding strictly positive   eigenfunction  $\phi _1$ is often called
  the ground state function.

   For the infinitesimal generator   $L_D$  of the semigroup we have  $L_D \phi _n(x)= -\lambda _n \phi _n(x)$
  The corresponding "heat" equation $\partial _t f(x,t)= L_D f(x,t)$  holds true  as well.

   It is useful to introduce the notion of the survival probability for the killed random process in a bounded domain  $D$, \cite{redner,killing}.
   Namely, given $T>0$, the probability that the random motion has not yet been absorbed (killed) and thus survives up to time $T$  is given by
   \be
   P^x[\tau >T] = P^x[X_T \in  D] = \int_D k_D(T,x,y) dy
\ee
and is  named  the survival probability up to time $T$.

Proceeding  formally  with  Eqs. (4) and (5), under suitable  integrability  and   convergence assumptions for the infinite series, we get:
\be
P^x[\tau >T]= \sum_{n=1}^{\infty } e^{-\lambda_nT}\, a_n \,  \phi _n(x) \,  \Rightarrow \,   a_1\,  e^{-\lambda_1T}\, \phi _1(x)
\ee
where $a_n = [\int_D  \phi _n(y) dy] $, $n=1,2,...$. We have arrived at  the  familiar exponential decay law  of the survival probability, characteristic for   e.g. the
 Brownian  motion with    absorbing boundary data, \cite{redner,killing}. Its time rate is  controlled by  the largest  eigenvalue  $ -\lambda _1$ of $\Delta _D$.
  Note that asymptotically the  functional  profile  ($x$-dependence)
   of the survival probability is kept stationary (exponential decay is executed as the continuous  scale change) and follows the pattern of  the
    eigenfunction $\phi _1(x)$.

\subsection{Conditioned random motions in a bounded domain}

For  the absorbing stochastic process with the transition density (4) (thus  surviving up  to time $T$), we  introduce
  survival probabilities  $P^y[\tau > T-t]$  and $P^x[\tau >T]$,  respectively  at times $T-t$ and $T$,  $0<t<T$.
 We infer a   conditioned  stochastic  process  with the transition density:
 \be
 q_D(t,x,y) = k_D(t,x,y) {\frac{P^y[\tau > T-t]}{P^x[\tau >T]}},
 \ee
 which   by construction   survives up to time  $T$  and  is additionally conditioned to start in  $x\in D$ at time $t=0$  and reach the
  target   point  $y\in D$, at time $t<T$.   An  alternative   construction  of such processes, in the diffusive case,  has been described  in \cite{killing},
   see also \cite{pinsky}.

 Given $t<T$, in the large time asymptotic of T, we can invoke (6), and  once  $T \rightarrow  \infty $ limit is executed, Eq.  (7) takes the form:
 \be
q_D(t,x,y) \longrightarrow   p_D(t,x,y) = k_D(t,x,y) {\frac{ \phi _1(y)}{ \phi _1(x)}}  \exp(\lambda _1 t)
 \ee
We have arrived at the  transition probability density  $p_D(t,xy)$  of the   probability conserving  process,  which never leaves the  bounded  domain $D$.
 Its   asymptotic (invariant) probability density   is $\rho (y) =[\phi _1(y)]^2$,  $\int_D \rho (y)\, dy =1$
  (that in view of  the implicit $L^2(D)$ normalization  of eigenfunctions $\phi _n$).

 By  employing (6)  and the definition  $\rho (y) =[\phi _1(y)]^2$,   we  readily check the stationarity property. We take  $\rho (x)$ as the initial
 distribution (probability density) of   points in  which the process is started  at time $t=0$).
 The  propagation towards target points,  to be reached at  time $t>0$,   induces a  distribution $\rho (y,t)$. Stationarity follows from:
\be
\rho (y,t)=  \int_D  \rho (x) p_D(t,x,y) dx = \rho (y).
\ee
Note that  in contrast to  $k_D(t,x,y)$ the transition probability function $p_D(t,x,y)$ is no longer a symmetric function of $x$ and $y$.

    \subsection{Quasi-stationary distributions}

In connection with so-called Yaglom limits.  \cite{bogdan1},  and in conjunction with the  previous description of the  conditioned random motions in a bounded domain,
 it is useful to   say few words  about the so called quasi-stationary distributions. These appear to be    a   useful  tool in the semi-phenomenological
analysis description of exponentially decaying in time populations,  whose probability distributions  display  specific   shape invariance on relatively long times scales,
while being close   extinction, see e.g.  \cite{collet,martinez}.      We borrow the idea, directly from   Ref. \cite{collet}.

Our major inputs are Eqs.(5)-(8).  Let us   define $\psi (x) = {\frac{1}{a_1}}  \phi _1(x)$ and   introduce  the expectation (mean)  value  of the function $f(x)$,
 with respect to $\psi (x)$,  as follows  \be
 \int_D \{\psi(x)  e^{\lambda _1t}\, E^x[f(X_t); t < \tau _D]\} dx   =
    \int_D   f(x)   \psi (x)dx = \int_D f(x) d\nu(x) = E_{\nu }[f]
\ee
We have  introduced a new  probability measure   $\nu $ on $D$   with   $\psi (x)$  as   its  probability  density. The latter  density stands for
  the  quasi-stationary distribution associated with the killed (absorbed) process in its large time regime,
c.f. Ref. \cite{collet}.

\subsection{Reflected motions in a bounded domain}

Reflected random motions in the bounded domain are typically expected to live indefinitely,  never leaving the domain,   basically with a complete reflection
form the boundary.  (We cannot a priori exclude  a partial reflection, that is accompanied by  killing or transmission.)

In case of previously considered motions a boundary  may be regarded as  either
a transfer terminal to the so-called "cemetry' (killing/absorption), or  as being  inaccessible  form the interior  at all  (conditioned processes).
In both scenarios, the major technical tool was the eigenfunction expansion (11), where the spectral solution for the Laplacian with the
 Dirichlet boundary data has been employed.  Thus,   in principle   we should here use the notation $\Delta _{\cal{D}}$, where
 ${\cal{D}}$ indicates  that the  Dirichlet boundary data   have been  imposed at the boundary  $\partial D$  of  $D\subset R^n$.

 Reflecting boundaries are related to Neumann boundary data, and then we shoul rather use the notation $\Delta _{\cal{N}}$.
 In a bounded domain we deal with  a spectral (eigenvalue) problem for $\Delta _{\cal{N}}$ with  the  Neumann data-respecting  eigenfunctions
 and eigenvalues.

  The major difference, if compared to the absorbing case is that the eigenvalue zero is  admissible and the corresponding
 eigenfunction  $\psi_0(x)$ determines an asymptotic (stationary, uniform in $D$) distribution  $\rho _0(x)= [\psi _0(x)]^2$, \cite{bickel0,bickel}.
In the Brownian context, the rough  form of the related  transition density   looks like:
\be
k_{\cal{N}} (t,x,y) = {\frac{1}{vol(D)}}  +  \sum_{n=1}^{\infty } e^{-\kappa_nt}\, \psi _n(x)\, \psi _n(y)
\ee
where   $\kappa _n$   are positive eigenvalues,  $\psi _n(x)$ respect the Neumann boundary data and $vol(D)    $  denotes
 the volume  of $D$ (interval length, surface are etc.).  We have $\psi_0(x)= 1/\sqrt{vol(D)}$.

\section{Fractional  Laplacians  in $R^n$}

In the present paper, up to    suitable  adjustment  of   dimensional constants,
     the free   evolution  in $R^n$  refers either to   $L= - \Delta$  (Brownian motion) or   $L= (-\Delta )^{\alpha /2}$ with $0<\alpha <2$
      (L\'{e}vy-stable motion).  It  is   $-(-\Delta )^{\alpha /2}$ which stands for  a  legitimate  fractional relative  of
      the ordinary Laplacian  $\Delta $.

For clarity of discussion let us recall three formal  (equivalent in $R^n$)  definitions of the
symmetric L\'{e}vy stable generator, which  nowadays
are  predominantly employed in the literature (we do not directly refer to  fractional derivatives).

The   spatially nonlocal fractional Laplacian has an  integral
definition (involving a  suitable function $f(x)$, with $x\in
R^n$)  in terms of the Cauchy principal value  (p.v.), that  is
 valid in space   dimensions $n\geq 1$
 \be
(-\Delta)^{\alpha /2}f(x)=\mathcal{A}_{\alpha,n} \lim\limits_{\varepsilon\to 0^+}
\int\limits_{{R}^n\supset \{|y-x|>\varepsilon\}}
\frac{f(x)-f(y)}{|x-y|^{\alpha +n}}dy. \ee
where  $dy \equiv d^ny$ and  the (normalisation) coefficient:
\be
\mathcal{A}_{\alpha,n}=
 \frac{2^{\alpha } \Gamma ({\frac{\alpha + n}{2}})}{\pi ^{n/2}
  |\Gamma (- {\frac{\alpha }{2}})|}  =
  \frac{2^{\alpha } \alpha \Gamma ({\frac{\alpha + n}{2}})}{{\pi ^{n/2}
  \Gamma (1- \alpha /2})}
\ee
Here one needs to employ   $\Gamma (1-s)= -s \Gamma (-s)$ for any $s\in (0,1)$.

has been adjusted  to secure that  the  integral   definition  stays in conformity   with its Fourier
transformed version. The latter  actually gives  rise to the
 Fourier multiplier representation  of the fractional Laplacian, \cite{kwasnicki,valdinoci,stef}:
\be
{\cal{F}} [(- \Delta )^{\alpha /2} f](k) = |k|^{\alpha } {\cal{F}} [f](k).
\ee
We recall again,  that it is $- (-\Delta )^{\alpha /2}$ which is a fractional analog of the Laplacian  $\Delta $.

We note that the formula (15) can be  rewritten in   the form, often exploited in the literature,
\cite{valdinoci,bucur}:
\be
(-\Delta)^{\alpha /2}f(x)=  {\frac{\mathcal{A}_{\alpha,n}}{2}} \int _{R^n}
\frac{2f(x) - f(x+y) -f(x-y)}{|y|^{n+\alpha }}\, dy.
\ee

Another definition, being  quite  popular in the literature  in view of the more explicit
dependence on the ordinary Laplacian, derives directly  from  the standard  Brownian semigroup  evolution  $\exp(t\Delta )$
The latter is explicitly  built into the formula, originally  related to   the  Bochner subordination  concept, \cite{kwasnicki}:
\be
(-\Delta )^{\alpha /2}f   = {\frac{1}{|\Gamma (-{\frac{\alpha }{2}})|}}\,
 \int_0^{\infty} (e^{t \Delta }f - f) t^{-1-\alpha /2}\, dt .
\ee
Clearly, given an initial datum $f(x)$, we deal here with   a solution of the  standard  (up to dimensional coefficient)  heat equation $f(x,t) = (e^{t \Delta }f$
  into the above  integral formula.

We note, that based on tools from functional analysis   (e.g. the spectral theorem), this definition of the fractional Laplacian
extends to   fractional powers of   more general  non-negative operators, than $(-\Delta )$ proper.

\section{Fractional Laplacians  in a bounded domain}

\subsection{Hypersingular (restricted)   fractional Laplacian}

  As mentioned before, a domain restriction to a bounded subset $D$ in $R^n$ is hard, if  not  impossible,  to implement via the Fourier multiplier definition.
The reason is an inherent spatial nonlocality of L\'{e}vy-stable generators.

Therefore, the natural way to handle e.g.  the  Dirichlet boundary data for a bounded domain $D$, one should begin from the hypersingular operator definition
(15)  and restrict its action to suitable  functions with  support in $D$. It is known that  the  standard Dirichlet restriction $f(x)= 0 $  for
 all $x\in \partial D$ is insufficient  for the pertinent functions. One needs to impose so-called exterior
  Dirichlet condition:   $f(x) = 0$ for all
 $x\in  R\backslash  D$.

By employing (15), (16) we  define  the  restricted   fractional Laplacian
 $(-\Delta )^{\alpha /2}_D$,  essentially   as  $(-\Delta )^{\alpha /2}$  of Eq. (15), with  a superimposed open domain $D$  restriction:
\be
(-\Delta )^{\alpha /2}_D f(x) = (-\Delta )^{\alpha /2}f(x)=g(x)
\ee
where   $x\in D$ and $f(x)=0=g(x)$ for all $x\in R^n\backslash D$. In particular, the spectral  (eigenvalue)  problem of interest
takes the form  $(-\Delta )^{\alpha /2}_D \phi (x)    = \lambda \, \phi (x)$. More detailed analysis of
 various eigenvalue problems  for  the  restricted fractional Laplacians   can be found  in Refs.  \cite{kulczycki}, \cite{duo}-\cite{zaba1} and
 \cite{lorinczi}-\cite{lorinczi2}.

 We note that Eq. (20)  can be converted to the form of the hypersingular Fredholm problem, discussed in detail in Refs.  \cite{zaba1,zaba2}.
 All involved singularities can be properly handled (are removable) and the pertinent formula reads:
$$
(-\Delta )^{\alpha /2}_D f(x) \equiv  - {\cal{A}}_{\alpha,n}\int _{\bar{D}} {\frac{f(u)}{|u-x|^{n + \alpha }}}\, du
$$

\subsection{Spectral fractional Laplacian}

 We first  impose the boundary conditions upon the Dirichlet  Laplacian in a bounded domain $D$  i. e. at the boundary $\partial  D$ of $D$.
 That is  encoded in the notation   $\Delta _{\cal{D}}$.  Presuming to have in hands its $L^2(D)$
  spectral solution (employed before in connection with (7)), we introduce a fractional power of the
  Dirichlet Laplacian $(- \Delta _{\cal{D}})^{\alpha /2}$ as follows:
  \be
  (-\Delta _{\cal{D}})^{\alpha /2}f(x) = \sum_{j=1}^{\infty } \lambda _j^{\alpha /2} f_j  \phi _j(x) =
  {\frac{1}{|\Gamma (-{\frac{\alpha }{2}})|}}\,
 \int_0^{\infty} (e^{t \Delta _{\cal{D}}}f - f) t^{-1-\alpha /2}\, dt .
  \ee
where  $f_j = \int _D f(x) \phi _j(x) dx$ and $\phi _j, j=1,2,...$ form an orthonormal basis
 system in $L^2(D)$:  $\int _D \phi _j(x) \phi _k(x) dx = \delta _{jk}$.

 We note that the spectral fractional Laplacian $(-\Delta _{\cal{D}})^{\alpha /2}$ and the  ordinary Dirichlet Laplacian $\Delta _{\cal{D}}$
 share eigenfunctions and their eigenvalues are  related as well: $\lambda _j \leftrightarrow \lambda _j^{\alpha /2}$.   The boundary data for
$(-\Delta _{\cal{D}})^{\alpha /2} $ are imported from these for  $\Delta _{\cal{D}}$.

 From the computational (computer-assisted)   point of view, this spectral simplicity has been considered as an advantage,
  compared to other proposals,  c.f. \cite{bonforte,vazquez}.

 In contrast  to the situation in $R^n$,  the restricted  $(- \Delta )^{\alpha /2}_D$   and spectral  $(- \Delta _{\cal{D}} )^{\alpha /2}$
   fractional Laplacians  are inequivalent  and  have  entirely  different sets of   eigenvalues and eigenfunctions.
 Basic differences between them have been studied in \cite{servadei}, see also \cite{valdinoci,bucur}  and \cite{duo}.

We note one most obvious (and not at all subtle) difference encoded in the very definitions:  the boundary data for the restricted  fractional  Laplacian need to be exterior
and set on $R^n\backslash D$, while those for the spectral one are set merely on the boundary $\partial D$  of $D$.

\subsection{Regional fractional Laplacian}

The regional fractional Laplacian has been introduced in conjunction with the notion of
censored symmetric  stable processes, \cite{bogdan,reflected}.
A censored stable process in an open set $D\subset  R^n$  is obtained from the
symmetric stable process by suppressing its jumps from  $D$  to the complement  $R^n \backslash D$
 of $D$, i.e., by restricting its L\'{e}vy measure to D.  Told otherwise,
 a censored stable process in an open   domain D is a stable process forced  to stay inside D.

Verbally that resembles random  processes  conditioned to stay in a bounded domain forever, \cite{killing}.
 However, we point out that the "censoring"  concept is not the same \cite{bogdan}  as that of the (Doob-type)
   conditioning  outlined.
   Instead, it is intimately   related to  reflected stable processes   in a bounded domain with killing within the domain, at its boundary
   and eventually not approaching  the boundary at all, \cite{bogdan,reflected}.

In Ref. \cite{reflected} the reflected stable processes in a bounded domain have been investigated, and their generators
identified with regional fractional Laplacians on the closed region $\bar{D}= D \cup \partial D$.
According to \cite{reflected}, censored stable processes  of Ref. \cite{bogdan},  in $D$ and  for
$0<\alpha \leq 1$, are essentially the same as the reflected stable process.  We shall  somewhat
undermine this view in below.

In general, \cite{bogdan}, if $\alpha geq 1$, the censored stable process will never approach $\partial D$. If
$\alpha >1$, the censored process may have a finite lifetime and may take values at $\partial D$.

Conditions for the existence of the regional Laplacian for all $x\in \bar{D}$
have been set in     Theorem 5.3 of \cite{reflected}.
For $1\leq \alpha <2$, the existence of the regional Laplacian  for all $x\in \partial D$, is granted if and only if
a derivative of a each function in the domain in the inward normal direction vanishes, \cite{reflected}.

For our present purposes we assume $0<\alpha <2$ and $D\subset  R^n$ being an open set.
The regional Laplacian is assumed to act upon functions $f$  on an open set $D$ such that
\be
\int _D {\frac{|f(x)|}{(1+ |x|)^{n+\alpha }}}  \,  dx < \infty
\ee
For such functions $f$, $x\in D$ and $\epsilon >0$, we  write
\be
(-\Delta )^{\alpha /2}_{D,Reg}  f(x)  = \mathcal{A}_{\alpha,n}
\lim\limits_{\varepsilon\to 0^+}
\int\limits_{y\in D \{|y-x|>\varepsilon\}}
\frac{f(x)-f(y)}{|x-y|^{\alpha +n}}dy.
\ee
provided the limit (actually the Cauchy  principal value, p.v.) exists.
Note a subtle difference between the restricted  and regional fractional Laplacians.
The former is restricted exclusively  by the domain property $f(x)=0, x\in R^n\backslash D$. The latter
is restricted by demanding the  integration variable $y$  of the L\'{e}vy measure  to be in $D$.

If we  superimpose (enforce)  the   (Dirichlet)  domain restriction upon the regional  fractional  operator  (for
a sufficiently  regular function   $f(x)$,   defined on the whole of $R^n$,   with the property $f(x)=0$ for
 $x\in R^n\backslash D$ of  an open set $D$),   we  arrive at the  identity, valid for all $x\in D$, \cite{bogdan}:
 \be
 (- \Delta )^{\alpha /2}f(x) -  (-\Delta )^{\alpha /2}_{D,Reg}  f(x)  = \kappa _D(x) f(x)
\ee
where:
\be
\kappa _D(x) =    \mathcal{A}_{\alpha,n}  \int_{R^n\backslash D} {\frac{1}{|x-y|^{n+\alpha }}} dy.
\ee
Note that Eqs. (23), (24), actually indicate how the restricted fractional Laplacian (20)  can be given  the  deeper  meaning.

We note that if to replace $D$   in Eq. (25)  by  $\bar{D}=D\cup \partial D$  one arrives at the definition of the   generator of  a
 reflected stable process  in $\bar{D}$, c.f. \cite{reflected},  $(-\Delta )^{\alpha /2}_{\bar{D},Reg} f(x) $, provided
  suitable conditions (various forms of  the  H\"{o}lder contituity) upon functions in the domain of the nonlocal operator are  respected.
In particular, in case of $1 \leq \alpha < 2$ it has been shown that $(-\Delta )^{\alpha /2}_{\bar{D},Reg} f(x) $
exists at  a boundary   point $x \in  ∂D$  if and only if   the normal inward derivative vanishes:
$(\partial f/\partial n)(x) = 0$.

\section{Random motion  in the interval}

 \subsection{Brownian motion}
 \subsubsection{Absorption  vs  conditioning  and quasi-stationary distributions}

Diffusion processes in the interval with various boundary  conditions  (Dirichlet, Neumann, mixed etc.)  have become favored model systems in the
 statistical physics  approach to the  Brownian motion,  including    extensions of the formalism to higher dimensions, \cite{redner,risken}. See also
 \cite{banuelos,banuelos1} for links with the previous formalism.

Let us consider the free   diffusion   (the customary  diffusion coefficient has been scaled away, e.g. set formally   $Dt \rightarrow t$)
  $\partial _t k=  \Delta _xk $  within an interval $(a,b) \subset  R$,
 with absorbing boundary conditions at its end points  $a$ and $b$.  Accordingly, we deal with the  Dirichlet  Laplacian  $\Delta _{\cal{D}}$.
 The   time  homogenous  transition density,   with $x, y  \in  (a,b)$, $0\leq s<t$ and $b-a=L$ reads

\be
k_{\cal{D}}(t,x,y) =  {\frac{2}L} \sum_{n=1}^{\infty } \sin\left( {\frac{n\pi
}L}(x - a)\right) \sin\left( {\frac{n \pi }L} (y - a)\right)  \exp
\left(- {\frac{n^2\pi ^2}{L^2}} \, t  \right).
\ee
Note  that  $\lim_{t\rightarrow s} k(x,t|y,s) \equiv \delta(x-y)$.

Let $c(x)$ be an arbitrary concentration function on the interval, $\int _D c(x)dx =0$.   Then $c(x,t)= \int_D k_{\cal{D}} (t,x,y) c(y) dy$  stands for
 a concentration at time $t>0$. Clearly $c(x,t)$ is a solution of the heat equation on the interval, e.g.   $\partial _t c(x,t) = \Delta _{\cal{D}}  c(x,t)$.

By employing the eigenfunction expansion (11)  we readily arrive at $c(x,t)= \sum _{n=1}^{\infty }  c_n e^{-\lambda _nt} \phi _n(x)$  with
$c_n = \int_D  \phi _n (y)  c(y) dy $.  Here: $\lambda _n = n^2\pi ^2/L^2$  and $\phi _n(x) =  \sqrt{2/L} \, \sin [(n\pi/L) (x - a)]$.

 The decay  of $c(x,t)$ for large times,   follows the exponential  pattern of Eq.  (13):
 \be
 c(x,t)  \longrightarrow  \sqrt{{\frac{2}L}} c_1 \sin\left( {\frac{\pi
}L}(x-a) \right)  \exp \left(- {\frac{\pi ^2}{L^2}} \, t  \right) =   c_1 \phi _1(x)  \exp (-\lambda _1 t).
  \ee
The survival probability is now slightly redefined  to the  form, \cite{redner},    $S(t) = \int_0^L c(x,t) dx$,
whose  large time asymptotic  reads $S(t) \sim c_1  \phi _1(x_0) \exp(-D\pi ^2  t/L^2) \equiv   c_1  \phi _1(x_0) \exp(-t/ \tau _0)$,  where
  $\tau _0 = 1/\lambda _1$ stands for the  decay time.

  For convenience, let us  note that  a transformation   $x\rightarrow  x'= (x-a)/L$   maps   the
interval $[a,b]$ into $[0,1]$.  Another transformation $x\rightarrow x'=x - {\frac{1}2} (a+b)$ maps  $[a,b]$ into $[-c,c]$, with $c=L/2$,
 whose special case (set $L=2$) is the interval $[-1,1]$.
With respect to the comparative  analysis of L\'{e}vy flights,  we  favor  the symmetric interval $[-c, c]$ with
$L=2c, c>0$.  (It is often convenient to make another scale change of the time parameter and ultimately   set  $D=1/2$).

 We mention the large time asymptotic of the transition density  (26):
\be
 k_{\cal{D}}(x,s;y,T)= k_{\cal{D}} (T-t,x,y)  \sim  \sin[{\frac{\pi} L}  (x+c)]\, \sin[{\frac{\pi } L} (y+c)] \, \exp
\left(- {\frac{\pi ^2}{L^2}} (T-s)\right)
 \ee
that is useful while evaluating (8) and (10).

The   emergent conditioned transition density (11) takes the form
\be
    p_D(t-s,x,y) =   k_D(t-s,x,y)  \, {\frac{\sin[{\frac{\pi} {2c}}  (x+c)]}{\sin[{\frac{\pi} {2c}}
(y+c)]}} \, \exp \left( +{\frac{\pi ^2}{4c^2}} (t-s)\right).
\ee
Note that  by construction  we  have $L=2c$ and   there holds
 $\sin[{\frac{\pi} {2c}}  (x+c)] = \cos({\frac{\pi} {2c}} x) $.

  By general principles  we  deduce  \cite{knight,pinsky}   the forward drift of
  the conditioned diffusion process  in question
    \be
  b(y) =  \nabla \ln \cos ({\frac{\pi}{2c}} y) =
  - {\frac{\pi}{2c}} \tan ({\frac{\pi}{2c}} y)
  \ee
 and the transport equation  for a probability density   in the    Fokker-Planck
 form (12)   (partial derivatives are executed with respect to $y$):  $ \partial _t  \rho = {\frac{1}2}  \Delta  \rho - \nabla (b\rho)$, with
  $\rho(y,t) = \int _{-c}^c \rho  (x) p_D(t,x,y) dx$.

  The asymptotic (invariant) probability distribution reads  (remembering about the $L^2(D)$
   normalization of the eigenfunctions: $\rho (x)   =  {\frac{2}L} \,   \cos^2(\pi x/L)$    and clearly refers to a diffusion process that is confined to stay in
    the interval forever (note a repulsion from the boundaries encoded in the drfit function).

In accordance with (13),   the  associated   quasi-stationary distribution   reads
$\psi (x)= (1/a_1)\phi _1(x)$ where $a_1= \int _D \phi (y) dy$.
In the present case we  have (the $L^1(D)$  normalization being implicit)
\be
\psi (x) = {\frac{\pi }{4c}}  cos ({\frac{\pi }{2c}} x)
\ee
which reads $(\pi /4)cos (\pi x /2)$, if  adapted to the interval $[-1,1]$, see e.g.  pp. 9 in  Ref. \cite{collet}

\subsubsection{Reflected Brownian motion}

The case of reflecting boundaries  in the interval   is specified by Neumann boundary conditions     for solutions of the diffusion  equation
$\partial _tf (x) = \Delta _{\cal{N}}f(x)$  in the interval $\bar{D}= [a,b]$.  We  need to have respected   $(\partial _x f) (a)= 0 =   (\partial _x f)(b)$ at
the interval boundaries.
  The pertinent transition density   reads:
  \be
k_{\cal{N}} (t,x,y)  =     {\frac{1}{L}}  +  {\frac{2}L} \sum_{n=1}^{\infty } \cos\left( {\frac{n\pi
}L}(x - a)\right) \cos\left( {\frac{n \pi }L} (y - a)\right)  \exp
\left(- {\frac{n^2\pi ^2}{L^2}} \, t  \right).
\ee

The operator $\Delta _{\cal{N}}$ admits the eigenvalue $0$ at the bottom of its spectrum, the corresponding eigenfunction being a constant. That refers to a
 uniform probability distribution on the interval of length $L$, to be approached in  the  asymptotic  (large time)  limit.
 Solutions of the diffusion equation with reflection at the boundaries of $D$  can be modeled  by setting $p(x,t)= k_{\cal{N}}(t,x,x_0)$, while remembering that $p(x,0)=
 \delta(x-x_0)$.  We can as well resort to  $c(x,t)= \int_D k_{\cal{N}} (t,x,y) c(y) dy$, while keeping in memory that $k(t,x,y)= k(t,y,x)$.

\subsection{L\'{e}vy flights}

\subsubsection{Restricted fractional case: hypersingular Fredholm problem}

In Refs.  \cite{zaba1,zaba3},  a reduction of the  definition (20) to the so-called hypersingular Fredholm problem
has been described.  let us choose $D=(-1,1) \subset R$.
 Essentially, under the  exterior Dirichlet boundary conditions, the fractional Laplacian $(-\Delta )^{\alpha /2}$, while acting on suitable functions that
 vanish everywhere on $R\backslash D$, acquires the form of the hypersingular operator (all potentially dangerous singularities are here removable, by a suitable
 regularization of integration, either in the sense of the Cauchy principal value or as the Hadamard-type regularization, \cite{zaba1}) :
 \be
 (-\Delta )^{\alpha /2}f(x) =  - {\cal{A}}_{\alpha }  \int_{-1}^1 {\frac{f(u)}{|u-x|^{1+\alpha } }}\, du
 \ee
 where ${\cal{A}}_{\alpha ,1} =  {\cal{A}}_{\alpha }= (1/\pi ) \Gamma (\alpha +1) \sin (\pi \alpha/2)$.  The  integral  needs to be understood  as the
   Cauchy   principal  value. relative to  $x\in (-1,1)$.    The eigenvalue problem for the operator  (33)  has been discussed in detail,  for various stability parameter values,
   with the  aid of numerical assistance, and compared with other existing solutions (analytic and computer assisted), see especially \cite{kwasnicki0,stos,zaba,duo}.

 Although analytic results are here scarse, we have a detailed knowledge of lowest eigenvalues and ground state functions shapes, that are relevant
 for the study of the large time asymptotic.
 The   validity of an  (approximate)  eigenvalue  formula for $n\geq 1$  and $0<\alpha <2$,
 \cite{kwasnicki0,stos}:
 \be
\lambda _n= \left[ {\frac{n\pi }{2}} - {\frac{(2-\alpha )\pi }{8}}\right]^{\alpha }  - O\left({\frac{2-\alpha }{n\sqrt{\alpha }}}\right)
\ee
has been extensively tested for the Cauchy case  ($\alpha =1$), with a number of partial observations concerning other   stability index values.
Let us emphasize that for $n\leq 10$, numerically computed eigenvalues are much sharper than these evaluated on the basis of Eq.  (34)  alone.
Thus e.g. in the Cauchy ($\alpha =1$ case the numerically computed bottom eigenvalues is
 $\lambda _1=1.157791$, while the leading part of the formula (34) would result in $\lambda _1= 1.178097$.

We note that the spectral solution for the ordinary  (minus) Laplacian in the interval
reads $\lambda _n= \left[ {\frac{n\pi }{2}}\right]^2$, $n\geq 1$.   Up to dimensional coefficients we have here the familiar quantum
mechanical spectrum of the infinite well se on the  interval  in question.

 In  Refs. \cite{duo,duo1}, in Table I,    a number of various eigenvalues for
  different stability indices has been  comparatively  collected.
Albeit with a rough accuracy, these data  give a quantitative picture of generic properties of the fractional Laplacian spectrum in restricted,
spectral and regional versions,  in the interval.

For the reader's convenience  we list lowest (ground state eigenvalues) for different stability indices: $\lambda (0.2)= 0.9575,  \lambda (0.5)=0.9702,
\lambda (0.7)= 1.1032, \lambda (0.9) =1,1032,
  \lambda (1) =1.1578, \lambda (1.2)= 1.2971, \lambda (1.5) =1.5976, \lambda (1.8) =2.0488, \lambda (1.95) =2.3520, \lambda (1.99) =2.4650$,
   to be set against the  bottom eigenvalue of the  standard  Laplacian  $(-\Delta _D)$:  $\lambda(2)= 2.4674$.

  Shapes of respective ground state eigenfunctions are not available in a closed analytic form and basic results in this connection
   (we leave aside the math-oriented research, \cite{kulczycki,kulczycki1,grzywny}) have been  obtained numerically,
     \cite{duo}-\cite{zaba3}, \cite{kwasnicki0}- \cite{zaba4}.

   Nonetheless, we can propose a general approximate formula ancompassing ground state functions for  all $0<\alpha <2$, whose accuracy has been extensively
    tested in the Cauchy case.  Namely, our proposal is to approximate $\phi _1(x)$ by
    \be
    \psi (x)= C_{\alpha ,\gamma } [(1-x^2) \cos(\gamma x)]^{\alpha /2},
    \ee
where $C_{\alpha ,\gamma } $ stands for the $L^2(D)$ normalisation factor, while $\gamma $ is considered to be the "best-fit" parameter,
allowing to get the best agreement with computer-assisted eigenfunction outcomes, \cite{zaba0}.

   In the Cauchy case, $\alpha =1$, almost prefect  fit (up to the available graphical resolution limit) has been obtained
   for  $\gamma  =  {\frac{1443}{4096}} \pi $, with $C= 0.92175$, \cite{zaba0}.

For the reader's convenience, we reproduce  a comparison of  rough approximations of few ground states with  the
 corresponding "best-fit" formulas.  These graphical outcomes have been obtained very recently, \cite{stefthanks}.

 The analytical expressions for approximate ground  functions, we compare with computer-assisted ground-state
  solutions  of the eigenvalue problems.

\begin{eqnarray}
&&\psi (x,\alpha =0.2)=0.786902\left[(1-x^2)\cos\frac{\pi x}{2}\right]^{0.1},\label{mu02}\\
&&\psi (x,\alpha =0.5)=0.876206\left[(1-x^2)\cos\frac{\pi x}{2}\right]^{0.25},\label{mu05}\\
&&\psi (x,\alpha =0.8)=0.90856\left[(1-x^2)\cos (1.3x)\right]^{0.4},\label{mu08}\\
&&\psi (x,\alpha =1.0)=0.921749\left[(1-x^2)\cos \frac{1443\pi}{4096}x\right]^{0.5},\label{mu1}\\
&&\psi (x,\alpha =1.5)=0.969531\left[(1-x^2)\cos (0.91x)\right]^{0.75}.\label{mu15}
\end{eqnarray}
The coefficients in the arguments of cosines have been chosen separately for
each $\alpha $  from the "best-fit"  assumption.

\begin{figure}[h]
\begin{center}
\centering
\includegraphics[width=95mm,height=95mm]{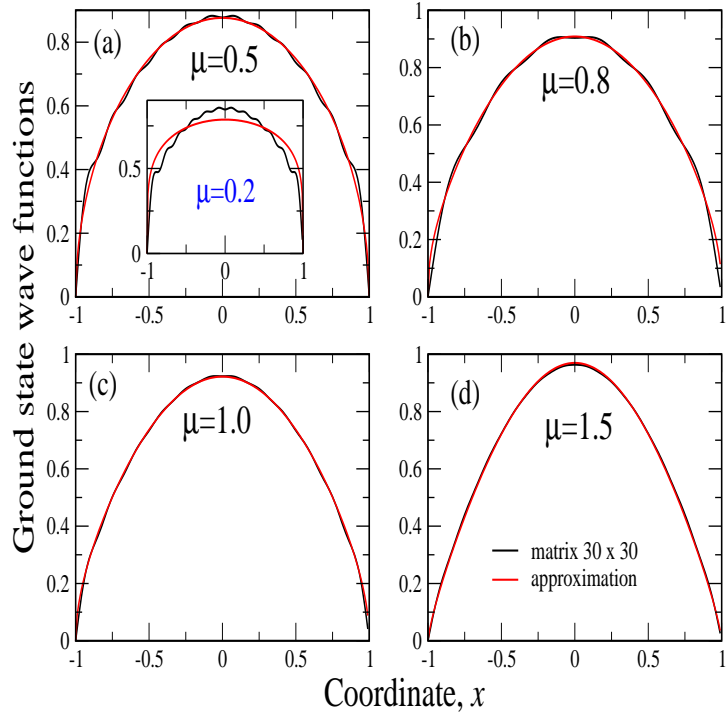}
\caption{Comparison of "exact" (for 30x30 matrix diagonalization, see \cite{zaba3})
 and approximate (Eqs. \eqref{mu02} - \eqref{mu15}) ground state  functions
  for different  stability  indices  $\mu$ (here our proviso is to use the notation  $\mu $
  instead of $\alpha $), shown in the panels. The inset to panel (a) reports the case of
$\mu=0.2$, when the approximation becomes  less  accurate than this  observed in
$\mu \geq 0.5$ regimes, \cite{stefthanks}. }
\end{center}
\end{figure}

{\bf Remark 1:}  The $(1-x^2)^{\alpha /2}$  behavior of the approximate ground state function   (35),
clearly conforms with results established in the mathematical literature, concerning the   near-boundary
properties of the involved "true" eigenfunction $\phi _1(x)$ corresponding to the bottom eigenvalue of
$(- \Delta )_D^{\alpha /2}$  (here, in the interval $[-1,1)$).
 Namely, it is known that for $x\in D$, we have a two-sided inequality
$$
c_1 \delta ^{\alpha /2 }(x)  \leq \phi _1(x) \leq c_2 \delta ^{\alpha /2 }(x)
$$
where $\delta (x)= dist(x,\partial D)$, while  constants $c_1, c_2$ depend on $D$ and the stability index
$\alpha $, see e.g. \cite{kulczycki,kulczycki1}. In the interval $(-1.1)$ that amounts to the comparability
criterion $\phi _1(x) \approx c_3\,  (1-x^2)^{\alpha /2}$,  where $c_3$ is a suitable constant.

{\bf Remark 2:}  The semigroup $T_t^{D}(\alpha ) = \exp(- t(-\Delta )^{\alpha /2}_D$, $t\geq 0$  of the stable process killed
upon exiting from a  bounded   set $D$ has an eigenfunction expansion of the form (7). Basically we never have in
hands a complete set of eigenvalues and eigenfunctions and likewise we generically do not know a closed analytic
form for the semigroup kernel $k_D(t,x,y)$, (7). A genuine mathematical achievement has been to establish
that when  $\alpha \in (0,2)$  and  a bounded domain $D$ is a subset of $R^n$, then  the stable semigroup
  $T_t^{D}(\alpha )$  is intrinsically ultracontractive. This technical (IU) property actually means that
  for any $t>0$ there exists  $c_t$ such that  for any $x,y \in D$  we have, \cite{kulczycki}:
  $$
k_D(t,x,y) \leq c_t\, \phi _1(x) \phi _1(y).
  $$
Actually we have $k_D(t,x,y) = \sum_1^{\infty } e^{-\lambda _nt} \phi _n(x) \phi _n(y)$.  Accordingly:
$$
 {\frac{k_D(t,x,y)}{e^{-\lambda _1 t} \phi _1(x) \phi _1(y)}}= 1 +
 \sum_2^{\infty } e^{-(\lambda _n- \lambda _1)t}
  \, {\frac{\phi _n(x) \phi _n(y)}{\phi _1(x) \phi _1(y)}}.
$$

It follows that we have a complete information about the (large time asymptotic) decay of
relevant quantities:
$$
lim_{t\rightarrow \infty }{\frac{k_D(t,x,y)}{e^{-\lambda _1 t} \phi _1(x) \phi _1(y)}} =1
$$
and  (for $t>1$)
$$
e^{-(\lambda _2- \lambda _1)t}  \leq  sup_{x,y \in D}  | {\frac{k_D(t,,x,y)}{e^{-\lambda _1 t}
\phi _1(x) \phi _1(y)}}|
\leq C_{\alpha ,D} e^{-(\lambda _2 - \lambda _1)t}.
$$
Thus, what we actually need to investigate the large time regime of L\'{e}vy processes in the
bounded domain $D$, is to know two lowest eigenvalues $\lambda _1, \lambda _2$  and the ground
state eigenfunction  $\phi _1(x)$ of the motion generator.

{\bf  Remark 3:}  The existence of conditioned L\'{e}vy  flights, with a transition density (11) and  an invariant
probability density $\rho (y) =[\phi _1(y)]^2$,  $\int_D \rho (y)\, dy =1$ is here granted as well.

\subsection{Spectral Dirichlet case}

In the bounded domain, the spectral definition (21) of the  Dirichlet fractional Laplacian, effectively reduces to
$(-\Delta _{\cal{D}})^{\alpha /2}f(x) = \sum_{j=1}^{\infty } \lambda _j^{\alpha /2} f_j  \phi _j(x)$, whose  eigenfunctions
are shared  with  the standard Dirichlet Laplacian $(-\Delta _{\cal{D}})$, while the corresponding eigenvalues are
raised to the power $\alpha /2$, e.g. read $\lambda _n^{\alpha /2}$, $n\geq 1$.  we emphasize that the boundary data refer to the boundary $\partial D$ of $D$
only.

In the context of  jump-type processes that are killed at the boundary, the spectral definition has been used explicitly in Ref. \cite{gitterman},  through
a direct analog of the  transition density (26):
\be
k^{\alpha }_{\cal{D}}(t,x,y) =  {\frac{2}L} \sum_{n=1}^{\infty } \sin\left( {\frac{n\pi
}L}(x - a)\right) \sin\left( {\frac{n \pi }L} (y - a)\right)  \exp
\left[ \left(- {\frac{n\pi}{L}} \right)^{\alpha } t\right] .
\ee
Here $0<\alpha <2$.
All elements of our discussion of asymptotic properties of the corresponding random motion, c.f. Sections II  and V remain
valid in the present spectral case.
In Ref. \cite{buldyrev}   a comparison has been made of the spectral and restricted Dirichlet definitions of fractional
Laplacians.  Numerical results for various average quantities do not substantially differ. It has been noticed that the restricted
Laplacian eigenfunctions  are close to the spectral Laplacian  eigenfunctions (trigonometric functions)
except for the vicinity of the boundaries.

However, in view of the spectral formula (34), the time rate formulas of the form (8),(9), (11), (27)
 and   those  listed in Remark 2 show up detectable differences.  It is also instructive to make a direct comparison of the
 pure Brownian case  (Section V.A) against the spectral one.

\subsection{Regional (censored vs reflected) L\'{e}vy flights}

Reflected L\'{e}vy flights in bounded domains as yet have not received a broad coverage in the literature,
\cite{bogdan,reflected} and the censored ones likewise.
Leaving the mathematical research thread somewhat aside, let us focus on interesting
findings in this connection,  in the
physics-oriented  publications, \cite{denisov,dubkov}.

Namely, in Ref. \cite{denisov} steady state  (stationary)   probability densities  for L\'{e}vy flights  in the interval $[-c,c]$, $L=2c$
(actually for an infinite well)  have been derived.

The departure point has been the standard fractional equation of the form (we scale
away all dimensional coefficients), c.f. Eq. (9) in Ref. \cite{denisov}:
\be
\partial _t f(x,t) = - (-\Delta )^{\alpha /2} f(x,t)
\ee
Infinitely deep potential well conditions are set in two steps,

 The first one amounts to a demand that
$f(x,t)=0$ for all $|x|>c$ while an interval of interest is [-1,1] and we say nothing specific about the values of
$f(x,t)$ at the boundary $\partial D$ of $D$. (The boundaries may be impenetrable, but the process make take values
on $\partial D$).

For the second step, we  invoke  the hypersingular integral formula (33), here  adapted to the interval $[-c,c]$
instead of the  original $[-1,1]$.

The stationarity condition is imposed in the form $\partial _t f(x,t)=0$, presuming that the
spectrum of the generator  (33) contains $0$ as the bottom eigenvalue  ((that in
  view of   $ (-\Delta )^{\alpha /2} \phi _n(x)= \lambda _n \phi _n$).
Hence,  we can formally write
\be
(-\Delta )^{\alpha /2} f(x) =  \int _{-c}^c {\frac{f(u) du}{|x-u|^{1+ \alpha  }}} =0.
\ee
The major assumption in Ref. \cite{denisov}  (by no means obvious and potentially
 questionable in view of hypersingular integral involved)  is that  Eq. (43) can be represented
 in the divergence form: $\\nabla j(x) =0$.
 It is an auxiliary condition, that the (formally) resulting  $j(x)$ vanishes everywhere in $[-c,c]$,
  from which there follows  the  $L^1[-c,c]$  normalized probability  density,  \cite{denisov},
in the closed  analytic form:
\be
\rho _{\alpha } (x)= (2c)^{1-\alpha }{\frac{\Gamma (\alpha )}{\Gamma ^2 (\alpha /2)}}
(c^2-x^2)^{\alpha /2-1},
\ee
 valid for  all  $0<\alpha \leq 2$.

The special case of the Cauchy noise ($\alpha =1$) has been addressed in Ref. \cite{dubkov},   by an
 independent reasoning,   with the outcome:
\be
\rho _{1} (x) = {\frac{1}{\pi }} {\frac{1}{\sqrt{c^2 -x^2}}}
\ee
valid for all $|x| < c$.

In passing we note that for $\alpha =2$ a uniform Brownian distribution $1/L$ arises. That would
suggest a link with reflected processes.

At the moment we cannot give an exhaustive analysis of affinities and/or differences between
 the censored and reflected L\'{e}vy processes.  As well we do not have a clear understanding
 whether the process,   associated with any   probability density
  $\rho _{\alpha }(x)$ given above, is or  is not  a reflected stable
   processes, which take values  at $\partial D$.

  In the whole stability parameter range  $0<\alpha <2$, the probability density
   $\rho _{\alpha }(x)$, Eq. (44),   blows up to infinity at the interval boundaries. Hence, the
    reflection condition of Ref. \cite{reflected} for $\alpha =1$
is manifestly violated: $\partial _x\rho _{1}(x)=   x/\pi (c^2-x^2)^{3/2}$ blows up to $\pm $ infinity at
 the interval boundaries, instead of  vanishing there.  This issue needs further analysis.

\section{Prospects}

We have described comparatively various aspects of random motion (Brownian and L\'{e}vy-stable),
contributing to the ongoing discussion (both from a a purely mathematical and more pragmatic, basically computer
assistance oriented, points of view). Definitely there is  some freedom in the definition of
 L\'{e}vy generators in a bounded domain that results in  giving access to new, not yet  exhaustively
  investigated, L\'{e}vy-type stochastic processes.  Their similarities and differences are
  surely worth an analysis as well. Additionally, some of the
 pertinent  definitions (specifically the spectral one)  have  gained popularity  in the study of
  nonlinear fractional problems (related to porous media), where they  have  proved to
  yield quite  efficient computer routines, see e.g. \cite{vazquez,bonforte}.

\end{document}